\documentclass[aps,prb,twocolumn,superscriptaddress,floatfix]{revtex4-1}
\usepackage[dvips]{graphicx,color}
\usepackage[table]{xcolor}
\usepackage[english]{babel}
\usepackage{colortbl}
\usepackage{color,enumerate,amsmath,amssymb,amsbsy,amscd,bm}

\newcommand{\ket}[1]{|{#1} \rangle}  
\newcommand{\bra}[1]{\langle {#1}|}  

\makeatletter
\def\CT@@do@color{%
\global\let\CT@do@color\relax
\@tempdima\wd\z@
\advance\@tempdima\@tempdimb
\advance\@tempdima\@tempdimc
\advance\@tempdimb\tabcolsep
\advance\@tempdimc\tabcolsep
\advance\@tempdima2\tabcolsep
\kern-\@tempdimb
\leaders\vrule
\hskip\@tempdima\@plus  1fill
\kern-\@tempdimc
\hskip-\wd\z@ \@plus -1fill }
\makeatother
\begin{document}
\title[]{Singularity of the time-energy uncertainty in adiabatic perturbation and cycloids on a Bloch sphere}
\author{Sangchul Oh}
\affiliation{Qatar Environment and Energy Research Institute, 
Hamad Bin Khalifa University, Qatar Foundation, PO Box 5825, Doha, Qatar}
\author{Xuedong Hu}
\affiliation{Department of Physics, University at Buffalo,
State University of New York, Buffalo, New York 14260-1500, USA}
\author{Franco Nori}
\affiliation{Center for Emergent Matter Science, RIKEN, Saitama 351-0198, Japan}
\address{Physics Department, The University of Michigan, Ann Arbor, Michigan, 48109-1040, USA}
\author{Sabre Kais}
\affiliation{Qatar Environment and Energy Research Institute,
Hamad Bin Khalifa University, Qatar Foundation, PO Box 5825, Doha, Qatar}
\affiliation{Department of Chemistry, Department of Physics and Birck Nanotechnology Center,
Purdue University, West Lafayette, IN 47907 USA}
\date{\today}
\begin{abstract}
The adiabatic perturbation is shown to be singular from the exact solution of a spin-1/2 particle
in a uniformly rotating magnetic field. Due to a non-adiabatic effect, its quantum trajectory on 
a Bloch sphere is a cycloid traced by a circle rolling along an adiabatic path. As the magnetic
field rotates more and more slowly, the time-energy uncertainty, proportional to the distance of
the quantum trajectory, calculated by the exact solution  is entirely different from
the one obtained by the adiabatic path traced by the instantaneous state. However,
the non-adiabatic Aharonov-Anandan geometric phase, measured by the area enclosed by
the exact path, approaches smoothly the adiabatic Berry phase, proportional to the area
enclosed by the adiabatic path. The singular limit of the time-energy uncertainty and
the regular limit of the geometric phase are associated with the arc length and arc area of
the cycloid on a Bloch sphere, respectively. Prolate and curtate cycloids are also traced by
different initial states outside and inside of the rolling circle, respectively.
The axis trajectory of the rolling circle, parallel to the adiabatic path, is shown to
be an example of transitionless driving. The non-adiabatic resonance is visualized
by the number of complete cycloid arcs.
\end{abstract}
\maketitle

%

Perturbation theory~\cite{Messiah,Bender} is widely used in many fields of science and engineering 
as an effective method to find an approximate solution to a given problem, expressed in terms of 
a power series in a small parameter. In regular perturbation calculations, one only keeps the first 
few terms of the expansion to obtain a good approximate solution to the exact one, as the small 
parameter goes to zero. However, there are many interesting problems that have no such uniform 
asymptotic expansion. These involve singular perturbations~\cite{wikisingular,Bender,Verhulst,
Johnson,Holmes,Shchepakina}. A classic example of this singular perturbation is the flow in the 
limit of zero viscosity~\cite{FeynmanLP}. When the viscosity, a small parameter, 
approaches zero, the solution of the Navier-Stokes equation gives a completely different solution 
from the one obtained by taking the zero viscosity to start out with.

Adiabatic perturbation~\cite{Messiah} is one of the fundamental approximations used in many fields.
Its classic applications include the Born-Oppenheimer approximation~\cite{Born-Oppenheimer} of 
decoupling for the fast electronic motion from the slow ionic one, and adiabatic quantum 
computation~\cite{Farhi}, an alternative to the quantum circuit model 
for quantum computing. The adiabatic theorem dictates that as long as a system changes slowly 
enough, a quantum system starting from an eigenstate would remain in the instantaneous eigenstate of the
time-dependent Hamiltonian up to the dynamical and Berry phases~\cite{Berry1984,Shapere}.
It may seem quite reasonable then that all physical properties in the adiabatic limit should be 
obtainable from the instantaneous eigenstate. However, this conjecture has never been proved.

In this paper, we reveal singular features of the adiabatic approximation by studying the quantum 
dynamics of a spin-1/2 particle in a uniformly rotating magnetic field. Its quantum trajectory 
is shown to be a cycloid on the Bloch sphere, traced by a point on a rolling circle, of a radius 
determined by the angular speed of the magnetic field, along the adiabatic path of the instantaneous 
eigenstate. We find the two basic geometric quantities, the distance and the enclosed area of 
the quantum trajectory, approach the different limits in the adiabatic limit. 
As the rotation of the magnetic field is slowed down, the non-adiabatic Aharonov-Anandan (AA) 
phase~\cite{AA1987}, the area enclosed by the quantum trajectory, goes to the adiabatic Berry phase,
the area enclosed by the adiabatic path. However, the time-energy uncertainty, the distance of the 
quantum trajectory, {\it does not} converge to the minimum time-energy uncertainty of the adiabatic path.
This singular feature of the adiabatic approximation is explained by the arc length and arc area
of the cycloid. In addition, the cycloid curve neatly explains some interesting physical results. 
First, the axis trajectory of a cycloid is interpreted as a transitionless driving that makes
the quantum evolution follow the adiabatic path. Second, the non-adiabatic resonance condition is
visualized by the number of perfect arcs of the cycloid. Finally, the exact cycloid, curtate and
prolate cycloids on a Bloch sphere are generated by different initial states.
Our results could be tested with a single qubit, a neutron, or light polarization, 
and could have important implications for the application of the adiabatic perturbation, for
example, adiabatic quantum computing and adiabatic quantum dynamics.
\medskip

\smallskip{\noindent\bf\sffamily Results\\[5pt]}
{\noindent\bf\sffamily\small Spin-1/2 particle in a rotating magnetic field.} We consider one of
the simplest quantum systems, a spin-1/2 particle in a rotating magnetic field
$\mathbf{B}(t) = B {\bf n}(t)$ where we assume its strength $B$ is constant and its direction
${\bf n}(t)$ rotates with constant angular speed $\omega$. Quantum dynamics is governed by
the time-dependent Schr\"odinger equation
\begin{align}
i\hbar\frac{d}{dpt}\ket{\psi} &= H(t) \ket{\psi}\,,
\label{Schrodinger_Eq}
\end{align}
where the time-dependent Hamiltonian is given by the Syman interaction
\begin{align}
H(t) = -\frac{\hbar\omega_0}{2}\,{\bf n}(t)\bm{\cdot\sigma}\,.
\label{Hamil}
\end{align}
Here $\omega_0$ is the Larmor frequency (for electron spin $\omega_0=egB/2m$), $\bm{\sigma}$
is the Pauli spin vector. As Feynman {\it et al.}~\cite{Feynman1957} showed, with a Bloch vector
${\mathbf r}(t) \equiv \bra{\psi(t)}\bm{\sigma}\ket{\psi(t)}$ Equation~(\ref{Schrodinger_Eq}) can be
written as the dynamics of spinning top
\begin{align}
\frac{d{\bf r}}{dt} = -\omega_0\,{\bf n}\times{\bf r}\,.
\end{align}

Before discussing the solution of Equation~(\ref{Schrodinger_Eq}), let us recall adiabatic
dynamics of a spin-1/2 particle. Adiabatic theorem dictates that
when an applied magnetic field changes slowly enough,
a quantum state $\ket{\psi(t)}$, adiabatically evolved from
an initial instantaneous state $\ket{{\bf n}_\pm(0)}$, remains in an instantaneous
eigenstate $\ket{{\bf n}_\pm(t)}$ up to the dynamical phase $\mp\frac{\omega_0}{2}t$ and
the adiabatic geometric phase $\gamma_\pm$ (called the Berry phase)~\cite{Berry1984,Shapere}
\begin{align}
\label{adiabatic_evolution}
\ket{\psi_{\rm ad}(t)} \simeq \exp\left(\pm i\tfrac{\omega_0}{2}t +i\gamma_\pm\right)\,\ket{{\bf n}_\pm(t)}\,.
\end{align}
The instantaneous eigenstates $\ket{{\bf n}_\pm(t)}$ are the solution of
$H(t)\ket{{\bf n}_\pm(t)} =\mp\frac{\hbar\omega_0}{2}\ket{{\bf n}_\pm(t)}$ and  written as
$\ket{{\bf n}_{+}(t)} =\cos\tfrac{\theta}{2}\,\ket{0} + e^{i\phi}\,\sin\tfrac{\theta}{2}\,\ket{1}
$ and $\ket{{\bf n}_{-}(t)} =-\sin\tfrac{\theta}{2}\,\ket{0} + e^{i\phi}\,\cos\tfrac{\theta}{2}\,\ket{1}
$. Here $\theta$ and $\phi$ are the polar and azimuthal angles of $\bf n$, respectively.
In adiabatic limit, the Bloch vector ${\bf r}(t)=\bra{\psi_{\rm ad}(t)}\pmb{\sigma}\ket{\psi_{\rm
ad}(t)}$, i.e., the spin direction is clearly align to the direction of the magnetic field ${\bf n}(t)$
in the parameter space.
The Berry phase is expressed in terms of the geometric quantity, the solid angle $\cal S$ subtended
by ${\bf n}(t)$ as $\gamma_\pm =\mp \frac{1}{2}{\cal S}$.
\smallskip

The question we would like to explore is how the non-adiabatic trajectory of the Bloch vector
${\bf r}$ approaches to that of $\bf n$ when the magnetic field rotates slowly. To this end,
the exact solution of Eq.~(\ref{Schrodinger_Eq}) is obtained by transforming it to the adiabatic
frame via the transformation $\ket{\psi(t)} = A(t) \ket{\varphi(t)}$
where $A(t)$ is composed of the column vectors $\ket{{\bf n}_{+}(t)}$ and $\ket{{\bf n}_{-}(t)}$.
In the adiabatic frame, the time-dependent Schr\"odinger Eq. becomes
\begin{align}
i\hbar\frac{\partial}{\partial t} \ket{\varphi(t)} = H_{\rm eff} \ket{\varphi(t)}\,,
\end{align}
where the effective Hamiltonian is decomposed into the sum of the adiabatic and the non-adiabatic
terms
\begin{equation}
\label{H_eff}
\begin{split}
H_{\rm eff}
&=  A^{-1} H A - i\hbar A^{-1}\frac{\partial A}{\partial t} \\
&= -\frac{\hbar\omega_0}{2}\sigma_z
+ \frac{\hbar\dot{\phi}}{2}(\mathbf{I} -\cos\theta\,\sigma_z + \sin\theta\,\sigma_x)
-\frac{\hbar\dot{\theta}}{2}\sigma_y\,.
\end{split}
\end{equation}
For the magnetic field rotating with constant angular speed, the effective Hamiltonian~(\ref{H_eff})
becomes time-independent and Eq.~(\ref{Schrodinger_Eq}) is exactly solvable.

We consider the two cases: (i) the rotation along the latitude,
$\phi = \omega t$ and constant $\theta$, and (ii) the rotation along the meridian,
$\theta = \omega t$ and constant $\phi$. Thus, in the adiabatic frame, the Bloch vector
${\bf r}(t)$ rotates with the frequency $\Omega$ around the new axis $\bf{\hat{e}}$ at acute
deviated angle $\alpha$ relative to the $\bf\hat{z}$ axis (adiabatic axis). The frequency $\Omega$
and the deviated angle $\alpha$ are given by $\Omega = \sqrt{\omega_0^2 + \omega^2}$
and $\alpha=\tan^{-1}\left({\omega}/{\omega_0}\right)$ for the rotation along the meridian,
and $\Omega = \sqrt{\omega_0^2 + 2\omega_0\omega\cos\theta + \omega^2}$ and $\alpha=\tan^{-1}
\left[{(\omega_0 + \omega\cos\theta)}/{\omega\sin\theta}\right]$ for the rotation
along the latitude, respectively.  The slowness of the rotation of the magnetic field is
relative to the Lamor frequency. So the ratio of two frequencies,
$\lambda\equiv \omega/\omega_0$ controls the adiabaticity of quantum dynamics.
In limit of $\lambda \to 0$, the quantum dynamics becomes adiabatic,
and the radius $a$ of the (imaginary) rolling circle along
the adiabatic path, given by $a=\sin\alpha$, becomes smaller.
The exact solution to Eq.~(\ref{Schrodinger_Eq}) is written as
\begin{align}
\ket{\psi(t)} = A(t)\, e^{i{\bm\sigma\cdot}{\bf{m}}\,\Omega t/2}\,
A(0)\ket{\psi(0)}\,.
\label{Eq_exact}
\end{align}

\begin{figure}[h]
\includegraphics[scale=1.0]{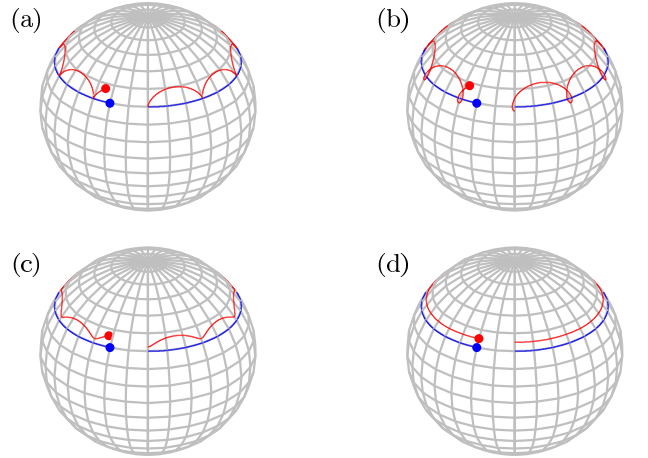}
\caption{{\bf\sffamily Cycloids on a Bloch sphere.} As the rotating magnetic field
traces the blue line, the Bloch vector makes various trajectories in red:
(a) the exact cycloid, (b) prolate cycloid, (c) curtate cycloid, and (d) the axis trajectory,
depending on the initial conditions.}
\label{Fig1}
\end{figure}

{\noindent\small\bf\sffamily Cycloids on a Bloch sphere.}
We calculate the trajectory of the Bloch vector ${\bf r}(t)$ on the Bloch sphere with the exact
solution Eq.~(\ref{Eq_exact}). Fig.~\ref{Fig1} plots the cycloids traced by the Bloch vector
${\bf r}(t)=\bra{\psi(t)}\pmb{\sigma}\ket{\psi(t)}$ on a Bloch sphere for different initial states
when the magnetic field is rotated clockwise with angular speed $\omega=0.5\,\omega_0$ around
the $z$-axis by azimuthal angle $\theta=\pi/3$. On a plane, a cycloid is the curve traced by
a point on the rim of a circle that rolls along a straight line. In classical mechanics, it is
the solution to two famous problems: brachistochrone (shortest-time) curves and tautochrone
(equal-time) curves~\cite{Goldstein,wikicycloid}. Like on plane, a cycloid on the Bloch sphere is traced by a point of
an imaginary circle which rolls along the base line. Here the adiabatic path
${\bf n}(t)$, i.e., the trajectory of the magnetic field, plays a role of the base line, as shown
by the blue curve in Fig.~\ref{Fig1}. The imaginary circle rolling along the adiabatic path
represents non-adiabatic quantum dynamics. The radius of the rolling circle is determined by two
frequencies $\omega_0$ and $\omega$, and given by $a = \sin\alpha$. Slower the rotation of
the magnetic field, smaller the rolling circle. This illustrates clearly how the quantum trajectory
approaches the adiabatic path in the adiabatic limit.

{\color{red}
}

\begin{table}[ht]
\setlength{\tabcolsep}{0pt}
\renewcommand{\arraystretch}{1.5}
\begin{tabular}{lcc}
\hline
\rowcolor{green!20}
& {\bf\sffamily plane cycloid} &{\bf\sffamily spherical cycloid}\\[5pt]
\hline
\textcolor{blue}{base line} & straight line  \quad & adiabatic path $\mathbf{n}(t)$\\[8pt]
\textcolor{blue}{circle radius}    & $a$
& $a = \sin\alpha$\\[8pt]
\textcolor{blue}{rolling speed} & $\varphi=\Omega t$
&$\Omega \equiv\sqrt{\omega_0^2 + \omega^2}$ \\[8pt]
\textcolor{blue}{equation}   &
$\left\{\begin{aligned}
x &= a\varphi -b\sin\varphi \\[2pt]
y &= a -b\sin\varphi
\end{aligned}\right.$
& $\begin{aligned}
   \textstyle
   i\tfrac{d\ket{\psi}}{dt}&= -\tfrac{\omega_0}{2}\,\mathbf{n}\cdot\bm{\sigma}\ket{\psi}\\[8pt]
   \quad\tfrac{d\mathbf{r}}{dt} &= -\omega_0\,\mathbf{n}\times\mathbf{r}
   \end{aligned}$\\[20pt]
\textcolor{blue}{curtate/prolate}     & in/outside & in/outside\\[8pt]
\textcolor{blue}{arc length} & $8a$       &$4a\cos\alpha\left[1+\frac{1-a^2}{2a}\ln\frac{1+a}{1-a}\right]$ \\[8pt]
\textcolor{blue}{arc area}   & $3\pi a^2$ & $2\pi a^2\left[1 + \frac{\cos^2\alpha}{1+\cos\alpha}\right]$
\\[5pt]\hline
\end{tabular}
\caption{ {\bf\sffamily Comparison between cycloids on a plane and on a sphere.} In plane case, the
curtate and prolate cycloids are traced by a point at radii $b < a$ and $b>a$, respectively.}
\label{table1}
\end{table}

\smallskip
Like a cycloid on plane, in addition to the instantaneous eigenstate of the initial Hamiltonian,
any initial states corresponding to a point on the rim of the rolling circle generates the cycloid.
Initial states inside and outside of the rolling circle trace the curtate cycloid and the prolate
cycloid on the Bloch sphere, respectively. The arc length and arc area of the cycloid on the sphere
are obtained by Bjelica~\cite{Bjelica2003}. Table~\ref{table1} shows the comparison of cycloids
on a plane and on a sphere. As shown by the blue curve in Fig.~\ref{Fig1}, the curve traced by the
axis of the rolling circle is parallel to the base line, i.e., the adiabatic path. The axis path can
be interpreted as an example of transitionless driving~\cite{Demirplak2003,Berry2009,Sangchul2014}
to accelerate the adiabatic evolution. This could be understood by considering the axis path as a new evolution
path of the spin, which is driven non-adiabatically by the parallel-rotating magnetic field, not
by a slowly rotating magnetic field along the axis path. Precisely, this can be done by adding
the transitionless-driving Hamiltonian $H_{D}(t)$ which cancels the non-adiabatic effect of
the original time-dependent Hamiltonian $H(t)$. So the new time-dependent Hamiltonian
$H(t) + H_D(t)$ drives non-adiabatically a quantum state along the adiabatic path of $H(t)$.
This technique could be used in flipping neutrons non-adiabatically by arrange flipping
In order to drive the spin-up state (north pole) to the spin-down state (south pole) along the
meridian line

\smallskip
{\noindent\bf\sffamily\small Non-adiabatic resonance.}
The non-adiabatic term, the second part in Hamiltonian~(\ref{H_eff}), causes the quantum trajectory
deviate from the adiabatic path. The question we address now is how the quantum evolution follows the
adiabatic path as the rotation of the magnetic field becomes slower. Even before approaching the
adiabatic limit, an evolved state could end up the adiabatic target state under some condition.
This is so called adiabatic resonance.
Let us consider the magnetic field is rotated by angle $\beta$
in time $T$, i.e., $\omega T= \beta$. Clearly, the Bloch vector ${\bf r}(t)$,
which is initially aligned to the $z$-axis in the adiabatic frame (that is, the instantaneous eigenstate
at $t=0$), will point again to the $z$-axis if a Bloch vector in the adiabatic frame is completely
rotated $n$ times, i.e., $\frac{\Omega T}{2} = 2\pi n$.
The non-adiabatic resonance condition is given by
\begin{align}
\frac{\Omega}{\omega} = \frac{4\pi}{\beta} n,
\end{align}
where $n$ is the number of cycloid arcs.

\smallskip
{\noindent\bf\sffamily\small Arc area of a cycloid as geometric phase}
Let us take a close look at physics associated with the geometric properties of a cycloid curve on
a Bloch sphere. The length and area are two basic geometric quantities of a curve. First, we discuss
the area enclosed by the cycloid curve. To this end, we consider again the magnetic field which
rotates completely around $z$-axis by azimuthal angle $\theta$ as shown in Fig.~\ref{Fig1}. As is well
known, when the magnetic field rotates slowly enough, the evolved state remains in the instantaneous
eigenstate, Eq.~(\ref{adiabatic_evolution}), and accumulates, in addition to the dynamical phase,
Berry phase which is proportional to the solid angle enclosed by the adiabatic
path ${\bf n}(t)$.~\cite{Berry1984} When the adiabatic condition is released, a cycled quantum state
accumulates the dynamical phase and the geometric phase, called Aharonov-Anandan (AA) phase, which
is proportional to the area of the curve of the quantum evolution in the projected Hilbert space~\cite{AA1987}.
When the resonance condition meets, the Bloch vector ${\bf r}(t)$ returns to its initial position
after a complete rotation of the magnetic field, so the cycloid curve is closed.
we calculate the AA phase and explore how the AA phase approaches to Berry phase in adiabatic limit.
With the exact solution~(\ref{Eq_exact}), AA phase is written as
\begin{subequations}
\begin{align}
\gamma_{AA} = \pi + \gamma_d
\end{align}
where
\begin{align}
\gamma_d = \frac{1}{\hbar}\int_0^T \bra{\psi(t)} H(t)\ket{\psi(t)}\ dt
= -\frac{\pi}{\lambda}\cos^2\alpha
\end{align}
\end{subequations}
In the adiabatic limit, i.e., $\lambda \to 0$, the dynamic phase $\gamma_d$ becomes
$\gamma_d \to -\pi\cos\theta$. Thus the AA phase converges to the Berry phase
\begin{align}
\lim_{\lambda\to 0} \gamma_{AA}[\lambda] = \gamma_{\rm Berry} = \pi(1-\cos\theta)\,.
\end{align}
As shown in Fig.~\ref{Fig1}, the difference between AA and Berry phases is the sum of the arc areas of
a cycloid. It is inversely proportional to running time $T$ so AA phase becomes Berry phase in
adiabatic limit as expected.

\smallskip
{\noindent\bf\sffamily\small Length of a cycloid as time-energy uncertainty and its singular limit}
Let us turn to physics related with the length of the cycloid curve. The Heisenberg position-momentum
uncertainty relation is based on the anti-commutation relation between two operators. However, the
energy-time uncertainty is different because time in quantum mechanics is not an operator.
Anandan and Aharonov~\cite{Anandan1990} gave a nice interpretation of the energy-time uncertainty
relation as the distance of the quantum evolution measured by the Fubini-Study metric in the
projective Hilbert space.
The length $L$ of the quantum evolution between two orthogonal
states, here from $\ket{0}$ to $\ket{1}$, is expressed as
\begin{align}
L = 2 \int_0^T \frac{\Delta E(t)}{\hbar}\,dt
=\frac{2}{\hbar}\langle \Delta E\rangle T,\quad
\end{align}
where $\Delta E = \sqrt{\bra{\psi} H^2 \ket{\psi} - \bra{\psi} H \ket{\psi}^2}$
is the uncertainty in energy during running time $T$. The length of any curve connecting two
orthogonal states $\ket{0}$ and $\ket{1}$ is greater than or equal to the shortest distance between
them, i.e., the geodesic line of the length $\pi$, $L\ge \pi$. So the energy-time uncertainty is given by
\begin{align}
\langle\Delta E\rangle T \ge \frac{h}{4}\,.
\end{align}
This tells that the minimum time $T$ required for transforming to an orthogonal state is
bounded by $T\ge h/4\langle \Delta E\rangle$, as shown by Mandelstam and Tamm~\cite{Mandelstam},
Fleming~\cite{Fleming}, Vaidman~\cite{Vaidman}, and Levitin and Toffoli~\cite{Levitin}.
\begin{figure}[h]
\includegraphics[scale=1.0]{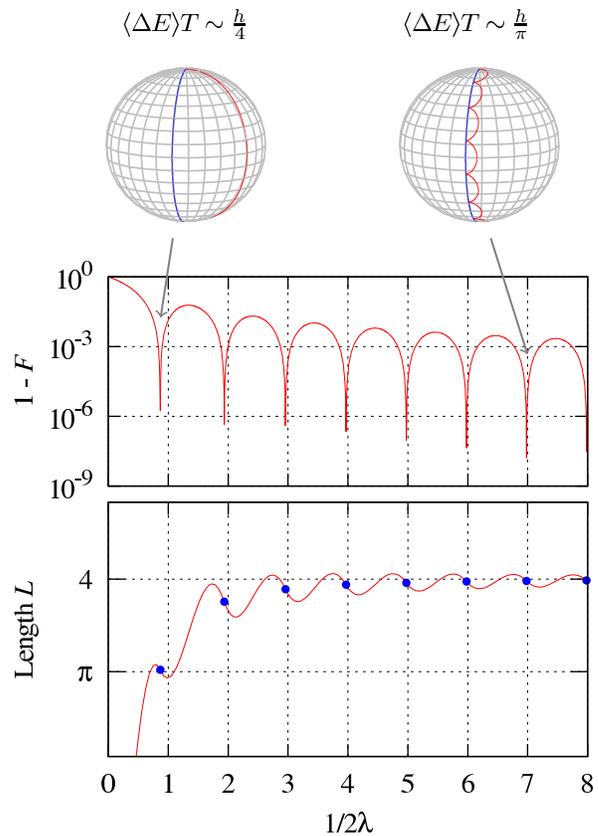}
\caption{{\bf\sffamily Trajectories, infidelity, and length.}
Top panel shows two trajectories in red on a Bloch sphere for $1/2\lambda=1$ and  7, respectively.
The blue line is the adiabatic path or the trajectory of a magnetic field.
Middle and bottom panels plot the infidelity, the probability deviating from $\ket{1}$,
and the distance of the quantum evolution, respectively, as a function of the adiabatic parameter
$\lambda$.  Blue points in bottom panel represent for the perfect transition.}
\label{Fig2}
\end{figure}

Now let us examine how the length of the quantum evolution changes as the speed of a rotating
magnetic field from the north to the south poles along the geodesic line varies, as shown in
Fig.~\ref{Fig2}. The adiabatic theorem dictates that if the magnetic field rotates slowly
enough, the quantum evolution is well approximated by an instantaneous eigenstate.
As depicted in Fig.~\ref{Fig2}, in adiabatic limit, the path of the quantum evolution approaches
to the adiabatic path, i.e., the geodesic line. One would expect that the length of the quantum
evolution in adiabatic limit becomes just that of the adiabatic path (geodesic line) because
the difference between the two paths, measured by the enclosed area (difference between AA and
Berry phases), becomes vanished. So the time-energy uncertainty in adiabatic limit would be minimum.
However, that is not the case. In the adiabatic limit the length of the quantum evolution becomes
$L=4$, but not $\pi$ as explained below. This is called the diagonal paradox~\cite{Weisstein}
or the limit paradox~\cite{Klymchuk} in calculus. Some well-known examples showing singular
limits~\cite{Berry2002} are the classical limit of quantum mechanics, and the limit of
zero viscosity~\cite{FeynmanLP} called d'Alembert's paradox.

With the exact solution of the Schr\"odinger equation, one can calculate the length of quantum
evolution. After some algebra, the length $L$ as a function of
adiabatic parameter $\lambda$ is given by
\begin{align}
  \begin{split}
L[\lambda] &=\frac{4}{\sqrt{1+\lambda^2}}\\
          &\times\int_0^{\frac{\pi}{2}\frac{\sqrt{1+\lambda^2}}{\lambda}}
  \sqrt{1 -\left[\cos^2x +\tfrac{1-\lambda^2}{1+\lambda^2}\,\sin^2x\right]^2}\,dx
\end{split}
\label{length}
\end{align}
In the limit of $\lambda \to 0$, while the integrand of Eq.~(\ref{length}) becomes smaller, the
interval of integration become larger. So one obtains
\begin{align}
\lim_{\lambda\to 0}L[\lambda] = 4\,
\end{align}
which is greater than the geodesic length $\pi$ between the north and the south poles.
This result can be also understood in terms of the product of the arc length of a cycloid
and the number of cycloid arcs needed. At non-adiabatic resonance condition with $\Omega T = 2\pi n$
and $\omega T = \pi$, the radius $a$ of the cycloid is given by
$ a = \frac{\omega}{\Omega} = \frac{1}{\sqrt{\lambda^2 +1}} = \frac{1}{2n}$.
In adiabatic limit, the cycloid can be seen as a plane cycloid, so the arc length is just $8a$.
Thus the length becomes $L \approx  8a\times n = 4.$

Fig.~\ref{Fig2} plots the infidelity, the probability of deviating from the target state $\ket{1}$,
and the distance $L$ of quantum evolution as function of the adiabatic parameter $1/\lambda=\omega_0/\omega$.
At the first non-adiabatic resonance, i.e., $1/2\lambda = 1$, the curve is composed of a single
cycloid and is deviated from the adiabatic path.  However, its length is lightly bigger than $\pi$.
This implies $\langle \Delta E\rangle T \approx h/4$. On the other hand, in adiabatic limit, i.e.,
$1/2\lambda \gg 1$, the curve is composed of many cycloids with smaller radius
and get close to the adiabatic path with length $\pi$. In adiabatic limit of $\lambda\to 0$,
while the red curve in Fig.~\ref{Fig2} is approaching to the blue one, the adiabatic path, its length
converges to $4$ not $\pi$, as shown before. Then the time-energy uncertainty becomes
$\langle \Delta E\rangle T \sim h/4$ which is not minimum. The quantum adiabatic theorem tells that
the instantaneous eigenstate~(\ref{adiabatic_evolution}) is a good approximation to the true quantum
evolution if the Hamiltonian changes slowly enough. While the non-adiabatic geometric phase
converges to the adiabatic geometric phase, the time-energy uncertainty doesn't.
This shows that {\it the adiabatic limit is singular} in the sense that the instantaneous
eigenstate cannot capture all physical properties in that limit.

\bigskip
{\noindent\small\bf\sffamily Discussion}\\

In conclusion we have shown that the Bloch vector of a spin in a rotating magnetic field traces
a cycloid on a Bloch sphere. Like on plane, different initial states trace prolate and curtate
cycloids, and the trajectory parallel to the adiabatic path. The perfect non-adiabatic resonance
is geometrically interpreted as the complete rolling of a cycloid. Two fundamental geometric
quantities, the area enclosed by a cycloid curve and its length, are connected to two physical
quantities, the geometric phase and the time-energy uncertainty, respectively. The arch areas
of a cycloid gives rise to the difference between AA phase and Berry
The energy-time uncertainty becomes $\langle \Delta E\rangle T \approx {h}/{\pi}$ which is
greater than the minimum energy-time uncertainty ${h}/{4}$.
We found the quantum adiabatic limit is singular, similar to the diagonal paradox or d'Amlembert's
paradox in the limit of zero viscosity. In adiabatic limit, while the AA phase converges to Berry
phase, the length, time-energy uncertainty, doesn't converge to that of the adiabatic path.

In mathematics, the isoperimetric inequality tells the relation between the circumference $L$ of a closed
curve and the area $A$ it encloses. The isoperimetric inequality on a sphere is obtained
by L\'evy~\cite{Osserman} and is given by
\begin{align}
L^2\ge A(4\pi -A)\,,
\end{align}
where the equality holds if and only if the curve is a circle. In our case, it tells the relation
between the geometric phase and the time-energy uncertainty. For a slow rotation of the magnetic field
along the great circle passing the north to the south poles, the area is given by $A=2\pi$, and
the quantum state acquires the phase factor, $-1$. If the quantum evolution were the great circle,
its length would be $L=2\pi$ according to the equality condition of isoperimetric inequality.
The distance of the quantum evolution is 8, but not $2\pi$, although it looks like a great circle.

The geometric phases, the area enclosed by a cycloid curve, have been measured with
various spin-$1/2$ systems such as a neutron in a rotating magnetic field, the polarization
of light in a coiled optical fiber, and qubits. The time-energy uncertainty, the length of
a cycloid curve, and its singular limit can be measured with same systems. While the geometric
phase is measured via interference between an evolved and initial quantum states,
a trajectory on a Bloch sphere seems to be needed for calculating the time-energy uncertainty
because of difficulty in measuring energy fluctuation. However, with the rapid advancement in
manipulating a qubit, it is possible to track a trajectory of a qubit on a Bloch sphere.
Especially, we notice that Roushan {\it et al.}~\cite{Roushan} traced the cycloid curve on a Bloch sphere
in the experiment of measuring the non-adiabatic geometric phase with superconducting qubits.

Adiabatic approximation is one of fundamental theorem in quantum mechanics, so it has many
applications, for example, Born-Oppenheimer approximation and adiabatic quantum computing.
The results here could give an opportunity to deepen our understanding of adiabatic approximation.

\bigskip
{\noindent\small\bf\sffamily Methods}\\
{\noindent\bf\sffamily Time-evolution of a spin in a rotating magnetic fields.}\\
The evolved quantum state at time $t$ is given by
\begin{align*}
\ket{\psi(t)} = A^\dag(t) U(t) A(0)\ket{\psi(0)}\,,
\end{align*}
where $U(t)$ is the time-evolution operator in the adiabatic frame.
For the magnetic field rotating about the $z$ axis by angle $\theta$, the effective Hamiltonian
in the adiabatic frame is given by
\begin{align*}
H_{\rm eff} &= -\frac{\hbar\omega_0}{2}\sigma_z -\frac{\hbar\omega}{2}\left(\cos\theta\sigma_z
-\sin\theta\sigma_x\right) \\
&= -\frac{\hbar\Omega}{2}\bf{m}\bm{\cdot\sigma}
\end{align*}
where $\Omega = \sqrt{\omega_0^2 +2\omega\omega_0\cos\theta + \omega^2}$,
${\bf m}=\cos\alpha\,\bm{\hat{z}} -\sin\alpha\,\bm{\hat{x}} =m_z \bm{\hat{z}} +m_x \bm{\hat{x}}$
is the direction of the effective magnetic field in the adiabatic frame, $m_z=\cos\alpha
=(\omega_0 +\omega\cos\theta)/\Omega$, and $m_x=-\sin\alpha=-\omega\sin\theta/\Omega$.
Thus the time-evolution in the adiabatic frame is given by
$e^{-iH_{\rm eff}t/\hbar} = e^{i\frac{\Omega t}{2}\bf{m}\bm{\cdot\sigma}}$.\\
\smallskip
{\noindent\small\bf\sffamily Calculation of time-energy uncertainty}
The magnetic field is rotated from the north pole to south pole along the geodesic line. The initial
state is $\ket{0}$. With the exact solution, it is straightforward to calculate the length
$L = \frac{2}{\hbar}\int_0^\tau \Delta E(t) dt$ with
$\Delta E(t) = \sqrt{\bra{\psi(t)}H^2(t)\ket{\psi(t)} -
\left(\bra{\psi(t)}H(t)\ket{\psi(t)}\right)^2}$.
One obtains $\bra{\psi(t)} H^2(t) \ket{\psi(t)} = {\hbar^2\omega_0^2}/{4}$
and
\begin{align*}
&(\bra{\psi(t)}H(t)\ket{\psi(t)})^2 \\
&= {\hbar^2\omega_0^2}/{4}\left[
\cos^2\left(\frac{\Omega t}{2}\right) +(m_z^2-m_x^2)\sin^2\left(\frac{\Omega t}{2}\right)\right]
\end{align*}
where  $m_z^2 -m_x^2=(\omega_0^2-\omega^2)/(\omega_0^2+\omega^2)= (1-\lambda^2)/(1+\lambda^2)$.
By changing the variable in the integrand, one obtains Equation~(\ref{length}).

\smallskip
{\noindent\small\bf\sffamily Calculation of time-energy uncertainty}
The trajectory of the Bloch vector $\bra{\psi(t)}\bm{\sigma}\ket{\psi(t)}$ is plotted with the exact
solution and by solving the time-dependent Schr\"odinger equation numerically with the Runge-Kutta
method.
\vfill


%
\bigskip
{\noindent\small\bfseries\sffamily Acknowledgments}\\
X.H. and S.O. were in part supported by the DARPA QuEST through
AFOSR and NSA/LPS through ARO.
\smallskip

\medskip
{\noindent\small\bfseries\sffamily Author contributions}\\
All authors, S.O., X.H., F.N., and S.K., analyzed and discussed the results and
wrote the manuscript together. S.O. conceived the research and
performed the calculation.

\medskip
{\noindent\small\bfseries\sffamily Additional information}\\
{\noindent\scriptsize\bfseries\sffamily Competing financial interests:} The authors declare no competing
financial interests.\\

\end{document}